\documentclass[prl,twocolumn,showpacs,superscriptaddress]{revtex4}
\usepackage{amsmath}
\usepackage{amssymb}
\usepackage{graphicx}
\usepackage{dcolumn}
\usepackage{bm}

\begin{document}
\title{
A Weighted Spike-Triggered Average of a Fluctuating Stimulus Yielding the Phase Response Curve
}
\author{Kaiichiro Ota}
\email{kaiichiro@acs.i.kyoto-u.ac.jp}
\affiliation{Graduate School of Informatics, Kyoto University, Yoshida-Honmachi, Sakyo-ku, Kyoto 606-8501, Japan}
\author{Masaki Nomura}
\affiliation{Graduate School of Letters, Kyoto University, Yoshida-Honmachi, Sakyo-ku, Kyoto 606-8501, Japan}
\author{Toshio Aoyagi}
\affiliation{Graduate School of Informatics, Kyoto University, Yoshida-Honmachi, Sakyo-ku, Kyoto 606-8501, Japan}

\date{\today}
\begin{abstract}
  We demonstrate that the phase response curve (PRC) can be
  reconstructed using a weighted spike-triggered average of
  an injected fluctuating input. The key idea 
  is to choose the weight to be proportional to the magnitude of the fluctuation 
  of the oscillatory period.  Particularly, when a neuron exhibits random switching
  behavior between two bursting modes, two corresponding PRCs
  can be simultaneously reconstructed, even from the data of a single trial.
  This method offers an efficient alternative to the experimental
  investigation of oscillatory systems, without the need for detailed modeling.
\end{abstract} 
\maketitle
\begin{figure}
  \includegraphics{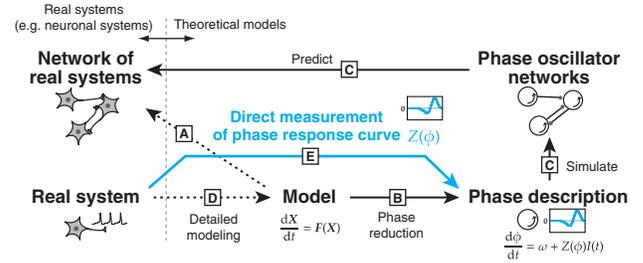}
  \caption{\label{concept}
  Without detailed modeling, the dynamical system reconstructed from the PRC obtained through 
  direct experimental measurements can be used to correctly predict the collective synchronization 
  properties of a set of coupled systems.
  } 
\end{figure}

Synchronization phenomena in networks consisting of interacting nonlinear
dynamical elements exhibiting limit-cycle oscillations have been the
subject of intensive study in physical, biological and social systems \cite{1-7}.
In general, such a system is described by ordinary differential 
equations of the form $\frac{\mathrm{d}\bm{X}}{\mathrm{d}t}=\bm{F}(\bm{X})$, 
where $\bm{X}$ denotes a multi-dimensional state of the system.
Owing to nonlinearity and complicated interactions,
however, we often encounter difficulties when attempting to analyze 
synchronization properties of such systems (see arrow A in Fig.\ref{concept}).
One of the most successful approaches for dealing with such difficulties is the phase
reduction method (arrow B), in which the evolution of each active
element is described by only one degree of freedom, the phase \cite{b7}.
With this method, the dynamics of $N$ coupled oscillatory systems can generally be described 
by a set of equations of the form 
$\frac{\mathrm{d}\phi_i}{\mathrm{d}t} = \omega_i+ \sum_{j=1}^N \Gamma_{ij}(\phi_i -\phi_j)$,
where $\omega_i$ is the natural frequency without coupling.
The phase of the $i$-th system representing the timing of its limit-cycle oscillation
is denoted by $\phi_i$.
This description is valid if the perturbation due to the interactions is so weak that
its effect only changes the phase asymptotically.  
The coupling function $\Gamma_{ij}(\phi)$ can then be derived from the original dynamical system as
$\Gamma_{ij}(\phi) \equiv \frac{1}{2\pi}\int_0^{2\pi} \bm{Z}_i(\theta)
\bm{V}_{ij}(\bm{X}^{(i)}_0(\theta),\bm{X}^{(j)}_0(\theta-\phi)) \mathrm{d}\theta$,
where $\bm{V}_{ij}(\bm{X}_i,\bm{X}_j)$ and $\bm{X}^{(i)}_0(\phi)$ represent the interaction exerted by the $j$-th 
system on the $i$-th system and the limit-cycle solution for the uncoupled $i$-th system, respectively \cite{b4}. 
Here, $\bm{Z}_i$ represents the phase response curve (PRC) of the $i$-th system. 
Therefore, if we can obtain the PRC, then we can predict the synchronization properties of 
the coupled system for any weak interaction (arrows C).

It is well known that the PRC can be calculated 
from the original model equations with the adjoint method \cite{38}.
Owing to technical difficulties limiting the applicability of experimental studies,
however, it is often difficult to construct 
system-specific detailed models including the essential properties for synchronization (arrow D).
For instance, if we wish to determine how a given neuromodulator causes
the synchronization properties of a coupled neuronal system to change, we cannot evaluate
all the effects of the neuromodulator on the dynamics of the various ion channels.
As an alternative approach, there have been some recent attempts to directly measure the PRC 
experimentally (arrow E).
The most important advantage of this approach is that, once the exact 
PRC is known, the dynamics of a set of weakly coupled oscillators are fully determined,
whether or not the original detailed dynamics are known. 
In the example considered above, therefore, all the effects of the neuromodulator are reflected 
in the measured PRC, and with this, the dynamical behaviors can be correctly predicted.

The standard method for obtaining PRCs through direct experimental measurements is to measure the phase shift 
induced by stimulating the system
with a brief pulse-like perturbation at various phases of the period \cite{impulse}.
Unfortunately, in such a study we often face the difficulty that the information from which we can derive the PRC
is lost in the noise arising from the uncontrollable nature of the environment,
though there are several statistical fitting methods which have been applied to efficiently extract the PRC from noisy data
\cite{PRCstat}.

An alternative approach for obtaining the PRC which we consider here, is to design a
different experimental procedure for the measurement, instead of the standard method described above.
In connection to this, G. B. Ermentrout \emph{et al.\ }have recently shown 
that the spike-triggered average (STA) of the injected noisy current 
is proportional to the derivative of the PRC under suitable conditions \cite{29}. 
This result inspired us to develop a more useful, novel method
for experimental protocols. 

\begin{figure}
  \includegraphics{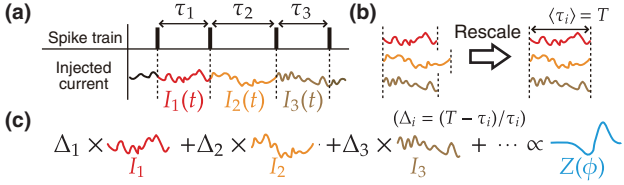}
  \caption{\label{method}
  Schematic of the experimental procedure to reconstruct the PRC in our method.
  (a) During the experiment, the
  injected fluctuating current $I(t)$ and the ISIs $\tau_i$ are recorded.
  The recorded $I(t)$ is divided into the current traces $I_i(t)$ 
  in the individual intervals. 
  (b) The rescaled currents $I_i(\tilde{t}_i)$ are obtained by adjusting the duration
  of each current trace $I_i(t)$ to the average $T$.
  (c) In the WSTA, the PRC can be reconstructed by averaging over all the rescaled currents 
  $I_i(\tilde{t}_i)$ with the weights $\Delta_i = \frac{T-\tau_i}{\tau_i}$, 
  the normalized difference between the ISI and the average interval.
  }
\end{figure}

In this paper, we demonstrate that the PRC can be 
directly reconstructed using an appropriately weighted spike-triggered average of the 
injected fluctuating inputs.
The key idea in this method is to choose the weight to be proportional to the magnitude of the fluctuation
of the oscillatory period. 
In the conventional method employing pulse-like perturbations, 
the spontaneous fluctuation of the period tends to make the estimation of the PRC quite rough.
Interestingly, the disadvantage of the fluctuating period in the case of the conventional method
becomes an advantage in our method. 
Although the proposed method can be applied to various real oscillatory systems,
such as Belousov-Zhabotinsky reactions, 
for simplicity, we consider two neuronal systems to explain the experimental procedure in the following.

{\it Experimental procedure -}
We consider the situation in which a neuron with some constant bias current 
exhibits spikes regularly with a period $T$. 
When an additional fluctuating current with zero mean is injected into this neuron,
the inter-spike interval (ISI) generally fluctuates about the average value $T$, as shown
in Fig.\ref{method}(a). 
As the first step of the experimental procedure, we record the stimulus current $I(t)$ and 
all the spike times over the entire spike train. 
Next, we divide the stimulus $I(t)$ into individual stimuli $I_i(t)$ 
between successive spikes, whose interval duration is denoted by $\tau_i$
($i=1,\dots,N$; $N$ denotes the number of the recorded spikes). 
Then (Fig.\ref{method}(b)), the slightly different duration times of the stimulus 
currents $I_i(t)$ are rescaled to the uniform period $T$ as
\begin{equation}
  \tilde{t}_i \equiv \frac{T}{\tau_i}t, \hspace{5mm}\tilde{t}_i \in [0,T].
  \label{eqn:rescale}
\end{equation}
Finally, the weighted spike-triggered average (WSTA) of the rescaled 
stimulus current $I_i(\tilde{t}_i)$ is defined as
\begin{equation}
  \mathrm{WSTA}(\tilde{t}_i) \equiv
  \left\langle \Delta_i I_i(\tilde{t}_i) \right\rangle,
  \label{eqn:wa}
\end{equation}
where $\Delta_i=\frac{T-\tau_i}{\tau_i}$, and
the angular brackets represent the average over all spikes, i.e.\ the index $i$ (Fig.\ref{method}(c)).
As shown below, we can prove that the above-defined WSTA is proportional to
the PRC, provided that the fluctuations in the stimulus current are not too strong. 
\begin{figure}
  \includegraphics{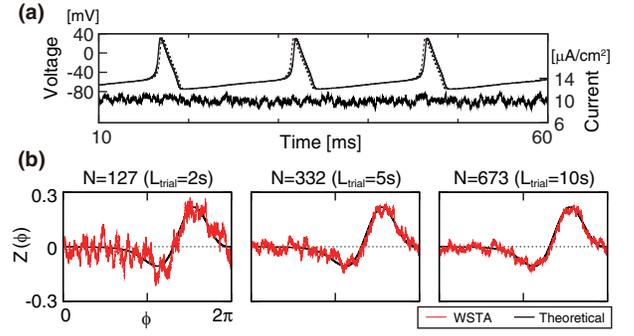}
  \caption{\label{HH}
  The estimated PRCs obtained from the WSTA in the case of the Hodgkin-Huxley model. 
  (a) 
  The time evolution of the membrane voltage (upper solid trace) 
  and the injected fluctuating current (lower trace).
  The injected current consists of two components:
  the bias current $I_{\mathrm{bias}}=10 \mathrm{\mu A/cm^2}$, and
  the zero-mean O-U noise with $\sigma=0.5 \mathrm{\mu A/cm^2}$.
  In the case that there exists only bias current, the neuron exhibits regularly periodic firing with a frequency
  of approximately 68Hz (upper dotted trace).
  (b) 
  The WSTA-estimated PRCs for three different measurement durations, $L_{\mathrm{trial}}$.
  The estimated PRC rapidly converges to the correct one as $L_{\mathrm{trial}}$ 
  increases. 
  }
\end{figure}

{\it Theory -}
When a neuron fires almost periodically under the influence of a zero-mean fluctuating stimulus current
$I(t)$, the corresponding phase dynamics can generally be described by the equation
\begin{equation}
  \frac{\mathrm{d}\phi}{\mathrm{d}t} = \omega + Z(\phi) I(t),
  \label{eqn:phase}
\end{equation}
in which  $\omega = \frac{2\pi}{T}$  and
$Z(\phi)$ represents the PRC with respect to the input current.
Let us consider the effect of the $i$-th stimulus, $I_i(t) \equiv \mu\xi_i(t)$, where
the parameter $\mu$ is introduced to account for the effect of the magnitude of the fluctuating 
current. Hereafter, we omit the subscript $i$ of the rescaled time $\tilde{t}_i$ 
for simplicity.  In addition, we assume that $\xi_i(t)$ is a stationary stochastic process with unit strength.
The auto-correlation of $\xi_i(t)$ is then defined as 
$C(\tilde{t})\equiv\left\langle  \xi_i(\tilde{t}) \xi_i(0) \right\rangle$. 
Then, substituting $I(t) = \mu \xi_i(t)$ into Eq.(\ref{eqn:phase}), 
and integrating from $t=0$ to $\tau_i$, we obtain
$\Delta_i = \frac{\mu}{2\pi} \int^{T}_0 Z(\phi_i(\tilde{s}))\xi_i(\tilde{s}) \mathrm{d}\tilde{s} \label{dtau}$,
where Eq.(\ref{eqn:rescale}) has been used.
Next, substituting this 
into Eq.(\ref{eqn:wa}), we find
\begin{equation}
  \mathrm{WSTA}(\tilde{t})
  = \frac{\mu^2}{2\pi} \int^T_0 \left\langle Z(\phi_i(\tilde{s})) \xi_i(\tilde{s}) \xi_i(\tilde{t}) \right\rangle \mathrm{d}\tilde{s}.
  \label{eqn:hat}
\end{equation}
We assume that we can expand the time development of the phase with respect to $\mu$ as
$\phi_i(\tilde{t}) = \omega\tilde{t} + \mu\phi_i^{(1)}(\tilde{t}) + \mu^2 \phi_i^{(2)}(\tilde{t}) + \cdots.$
Using this and the Taylor expansion in Eq.(\ref{eqn:hat}), we get 
\begin{align}
  &\mathrm{WSTA}(\tilde{t}) \nonumber \\
  &= \frac{\mu^2}{2\pi} \int^T_0 \left\langle \left\{ Z(\omega\tilde{s}) + \frac{\mathrm{d}Z}{\mathrm{d}\phi} \omega\mu\phi_i^{(1)} + \cdots \right\} \xi_i(\tilde{s}) \xi_i(\tilde{t}) \right\rangle \mathrm{d}\tilde{s} \nonumber \\
  &= \frac{\mu^2}{2\pi} \int^T_0 Z(\omega\tilde{s}) C(\tilde{s}-\tilde{t})
  \mathrm{d}\tilde{s} + O(\mu^3).
\end{align}
Assuming that the timescale on which the auto-correlation of the stimulus current decays is much shorter than
the period $T$, we can approximate $C(\tilde{t})$ as $\delta(\tilde{t})$, where $\delta$ is the Dirac delta function.
Using the fact that $\mu$ is then given by $\left\langle I_i(\tilde{t}) I_i(0) \right\rangle = \mu^2 \delta(\tilde{t})$, we finally obtain 
\begin{equation}
  \mathrm{WSTA}(\tilde{t})= \frac{\mu^2}{2\pi} Z\left(\phi\right) + O(\mu^3),
\end{equation}
where $\phi=\omega\tilde{t}$.
We have thus found that the PRC can be reconstructed from the spike-triggered average 
of the stimulus current using a weight 
proportional to the normalized difference between the ISI and the average interval.

In the numerical examples considered below, to generate fluctuating stimuli,
we used the zero-mean Ornstein-Uhlenbeck (O-U) process 
$\mathrm{d}I_t = -\gamma I \, \mathrm{d}t + \sqrt{2\gamma \sigma^2} \, \mathrm{d}W_t$, where $W_t$ is a Wiener process.
For sufficiently large $\gamma$ and in the long-time limit, we can approximate 
$\left\langle I_t I_0 \right\rangle$ as $\sigma^2 e^{-\gamma |t|} 
\cong (\int^\infty_{-\infty}\sigma^2 e^{-\gamma |t|} \mathrm{d}t) \delta(t) = \frac{2\sigma^2}{\gamma} \delta(t)$
(i.e. $\mu=\sqrt{2/\gamma}\,\sigma$).
In all the examples considered here, we fixed $\gamma = 5$ and varied $\sigma$ as the control parameter.

\begin{figure}
  \includegraphics{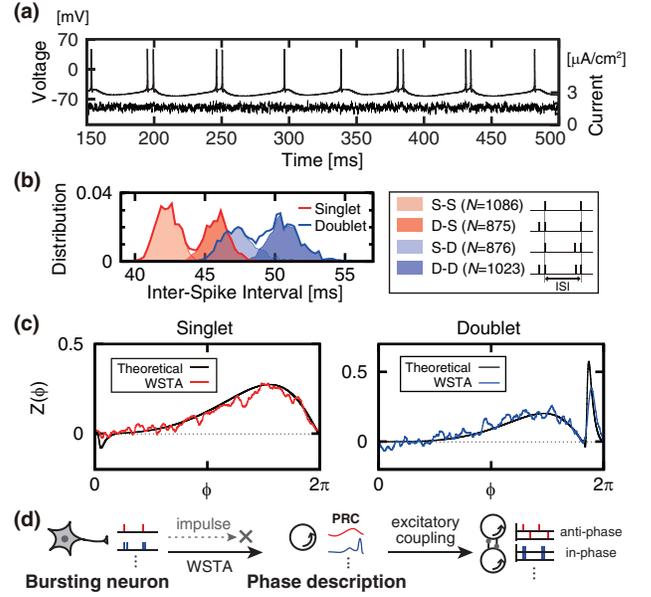}
  \caption{\label{CH}
  From a single fluctuating current ($L_{\mathrm{trial}}=$180 sec) that 
  causes a bursting neuron to exhibit irregular singlet or doublet firing, two PRCs 
  corresponding to the singlet and doublet states can be reconstructed simultaneously. 
  (a)
  The membrane voltage of the bursting neuron (upper trace) and the injected current (lower trace,
  $\sigma=0.2 \mathrm{\mu A/cm^2}$ and $I_{\mathrm{bias}}=1.605 \mathrm{\mu A/cm^2}$).
  (b) 
  Distributions of conditional ISIs. The multi-modal distribution consists of four unimodal
  distributions characterized by two successive firing modes;
  for example, S-D denotes the distribution of the ISI between the initial {\it singlet}
  and the final {\it doublet} firings. 
  (c)
  Comparisons between the WSTA-estimated PRCs and the theoretical ones.
(d)
  A distinct difference between two WSTA-estimated PRCs 
  leads to a transition in the synchronization for a two-neuron system.
  }
\end{figure}

{\it Numerical examples -}
To confirm the validity of our proposed method, we compare the PRCs obtained from the WSTAs
and the PRCs calculated using the adjoint method.
In Fig.\ref{HH}, we first show typical results 
in the case of the Hodgkin-Huxley model \cite{20}. Figure \ref{HH}(a) displays a typical voltage 
trajectory in response to the fluctuating current.
Figure \ref{HH}(b) presents comparisons between the true 
PRC (black trace) and the PRC estimated from the WSTA (red trace) for three different 
measurement durations. We see that as the measurement duration 
increases, the WSTA-estimated PRC converges to the true 
one.  This result suggests that the PRCs of real systems can be accurately estimated within 
practically reasonable recording times experimentally.

We next consider the more complicated situation in which a bursting neuron exhibits singlet 
or doublet firing randomly, under the influence of the injected fluctuating current, as shown in Fig.\ref{CH}(a).
For the bursting neuron, here we adopt the chattering neuron model, which exhibits an increasing 
number of intra-burst spikes, such as from singlet to doublet, when the injected current is increased \cite{41}.
In the case considered in Fig.\ref{CH}, the injected current fluctuates about the mean level
corresponding to the transition between singlet and doublet firings.
Using the WSTA with an additional procedure,
we can simultaneously reconstruct two PRCs corresponding to singlet and 
doublet states from a single trial.
The key idea here is to separately calculate two conditional WSTAs,
depending on whether a singlet or doublet firing occurs.
In other words, the PRC for the singlet (doublet) firing can be reconstructed 
by calculating the WSTA over only the injected currents generating the singlet (doublet) firing. 
In Fig.\ref{CH}(b), closer investigation reveals that 
the multi-modal distribution of the ISI can be regarded as a mixture of four different unimodal 
distributions, which are specified by two successive firing modes.
From each unimodal distribution, the average ISI is separately computed for calculation 
of the $\Delta_i$ in Eq.(4). Finally, two rescaled sets of data with the same final firing mode are used
to obtain the two PRCs for the singlet and doublet modes.
Figure \ref{CH}(c) demonstrates that both of the PRCs obtained in this way are in reasonably good agreement with the exact ones. 
We emphasize that two different PRCs can be reconstructed from a single-trial data set, 
even when the naive conventional method employing a pulse-like perturbation is not practical.
Noting the sharp difference between these two WSTA-estimated PRCs,
we predict a transition between anti-synchronous and
synchronous firing for a two-neuron system with excitatory synaptic couplings, 
as shown in a previous study \cite{31}(Fig.\ref{CH}(d)).

\begin{figure}
  \includegraphics{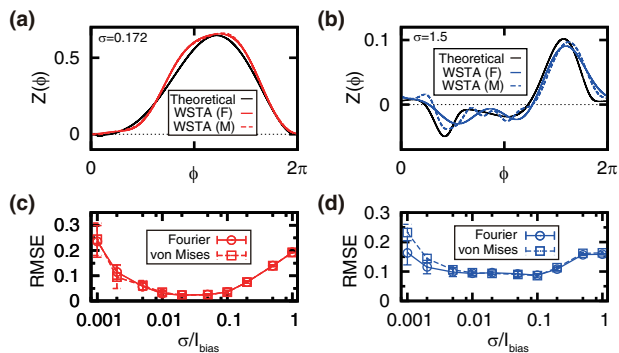}
  \caption{\label{ML}
  The optimal strengths of the injected fluctuation in a noisy environment.
  As the background noise, zero-mean O-U noise with $\sigma=0.001$ 
  was added to the dynamics of the potassium activation variable. 
  In all cases, the WSTAs were calculated over 1000 spikes.
  (a-b) The estimated PRCs for two types of Morris-Lecar models:
  (a) type I excitability near the saddle-node bifurcation point ($I_{\mathrm{bias}}=8.6 \mathrm{\mu A/cm^2}$, 
  $T^{-1}=$28Hz);
  (b) type II excitability near the Hopf bifurcation point ($I_{\mathrm{bias}}=30 \mathrm{\mu A/cm^2}$, 
  $T^{-1}=$64Hz).
  For smoothing algorithms, we considered several statistical regression models, 
  in which 
  Fourier (F) and von Mises (M) functions were used as basis functions \cite{regression}.
  (c) The dependence of the root-mean-square error (RMSE) 
  on 
  $\sigma/I_{\mathrm{bias}}$ for the type I model (5 trials).
  (d) The same graph as in (c) for the type II model.
}
\end{figure}

{\it Discussion -}
In actual applications to real systems, we have to choose an appropriate magnitude of 
the fluctuation of the injected current. 
Since some unpredictable and uncontrollable 
noise is inevitable in real systems, we examine the discrepancy between the estimated PRC and the true one
in the presence of unknown background noise, as summarized in Fig. \ref{ML}.
We find that for two types of Morris-Lecar models near the threshold for firing \cite{ML},
the WSTA estimates the PRC most precisely in the case that an intermediate magnitude of the fluctuation of
the injected current is chosen. 
This is because the signal of the PRC becomes lost in the background noise if
the injected fluctuations are too weak, while if the fluctuations are too strong, 
the approximation of the phase description becomes poor. 
Figures \ref{ML}(c) and (d) suggest that the same results hold 
for all types of bifurcations and smoothing algorithms considered.

We now give some comments on three relevant studies.
\marginpar{c1}First, when the fluctuation of the ISIs vanishes, all the weights of the WSTA are the same.
This is equivalent to the situation in which the STA yields $Z^\prime(\phi)$ instead of $Z(\phi)$ \cite{29}.
This superficial inconsistency can be resolved by considering the fact that the difference between the WSTA and 
the STA converges to zero as $\mu^2$.
\marginpar{c1}Second, recent studies have pointed out that
a correction term appears in the phase description (\ref{eqn:phase})
if the limit-cycle oscillation is perturbed by the noise with a correlation time shorter
than the relaxation time of the limit-cycle attractor\cite{Y-term}.
\marginpar{c2}Under real conditions, a fluctuating noisy signal can be regarded as a smooth signal in the limit of
a small time scale. Therefore, Eq.~(\ref{eqn:phase}) and our result are practically valid.
\marginpar{c1}Third, in another study\cite{kawamura2008}, 
the collective PRC to a macroscopic external force for globally coupld oscillators is investigated. 
\marginpar{c2}In principle, our method can be applied to measure such a collective PRC.
\marginpar{c3}Closer examinations of the above-mentioned issues 
are beyond the scope of this paper, but they are of great importance and should be studied further.

In conclusion, we demonstrated that the PRC can be reconstructed from
our proposed WSTA, in which the weight is proportional to the
magnitude of the fluctuation of the period.
In this study, considering limit-cycle oscillations observed ubiquitously in nonlinear dissipative systems far from
equilibrium, we found a theoretical relation between 
the fluctuations in the system and its response to an external force.
This might provide useful insight for further studies as the development of fluctuation dissipation theorem.
In comparison with the standard method employing pulse-like perturbations,
furthermore, the proposed WSTA method is experimentally reliable, fast and wide applicable,
as shown by the fact that two different PRCs can be reproduced even from a single-trial data set 
for two-mode bursting neurons.
We believe that this method will contribute to elucidating the nature of
real dynamical system in a broad range of contexts.
\begin{acknowledgments}
  This work was supported by KAKENHI(20033012, 18300079, 19GS2008) from MEXT, Japan.
\end{acknowledgments}

\end{document}